\documentstyle[twoside]{article}

\catcode`\@=11
\long\def\@makefntext#1{
\protect\noindent \hbox to 3.2pt {\hskip-.9pt  
$^{{\eightrm\@thefnmark}}$\hfil}#1\hfill}		%CAN BE USED 

\def\@makefnmark{\hbox to 0pt{$^{\@thefnmark}$\hss}}	%ORIGINAL 
	
\def\ps@myheadings{\let\@mkboth\@gobbletwo
\def\@oddhead{\hbox{}
\rightmark\hfil\eightrm\thepage}   
\def\@oddfoot{}\def\@evenhead{\eightrm\thepage\hfil
\leftmark\hbox{}}\def\@evenfoot{}
\def\sectionmark##1{}\def\subsectionmark##1{}}

\oddsidemargin=\evensidemargin
\newcounter{sectionc}\newcounter{subsectionc}\newcounter{subsubsectionc}
\renewcommand{\section}[1] {\vspace{12pt}\addtocounter{sectionc}{1} 
\setcounter{subsectionc}{0}\setcounter{subsubsectionc}{0}\noindent 
	{\tenbf\thesectionc. #1}\par\vspace{5pt}}
\renewcommand{\subsection}[1] {\vspace{12pt}\addtocounter{subsectionc}{1} 
	\setcounter{subsubsectionc}{0}\noindent 
	{\bf\thesectionc.\thesubsectionc. {\kern1pt \bfit #1}}\par\vspace{5pt}}
\renewcommand{\subsubsection}[1] {\vspace{12pt}\addtocounter{subsubsectionc}{1}
	\noindent{\tenrm\thesectionc.\thesubsectionc.\thesubsubsectionc.
	{\kern1pt \tenit #1}}\par\vspace{5pt}}
\newcommand{\nonumsection}[1] {\vspace{12pt}\noindent{\tenbf #1}
	\par\vspace{5pt}}

\newcommand{\textlineskip}{\baselineskip=13pt}
\newcommand{\smalllineskip}{\baselineskip=10pt}

\def\eightcirc{
\begin{picture}(0,0)
\put(4.4,1.8){\circle{6.5}}
\end{picture}}
\def\eightcopyright{\eightcirc\kern2.7pt\hbox{\eightrm c}} 

\newcommand{\copyrightheading}[1]
	{\vspace*{-2.5cm}\smalllineskip{\flushleft
        {\footnotesize {\bf Gravitation and Cosmology vol. 5 
         No. 2(18) (June 1999) 81-91} #1}\\
        {\footnotesize $\eightcopyright$\, 1999 by H.C. Rosu: gr-qc/9412012 v4
        }\\
	 }}

\def\abstracts#1#2#3{{
	\centering{\begin{minipage}{4.5in}\baselineskip=10pt\footnotesize
	\parindent=0pt #1\par 
	\parindent=15pt #2\par
	\parindent=15pt #3
	\end{minipage}}\par}} 

\newcommand{\bibit}{\nineit}

\renewenvironment{thebibliography}[1]
	{\frenchspacing
	 \ninerm\baselineskip=11pt
	 \begin{list}{\arabic{enumi}.}
        {\usecounter{enumi}\setlength{\parsep}{0pt}     
	 \setlength{\leftmargin 12.7pt}{\rightmargin 0pt} %FOR 1--9 ITEMS
         \setlength{\itemsep}{0pt} \settowidth
	{\labelwidth}{#1.}\sloppy}}{\end{list}}

\newcounter{itemlistc}
\newcounter{romanlistc}
\newcounter{alphlistc}
\newcounter{arabiclistc}

\def\@citex[#1]#2{\if@filesw\immediate\write\@auxout
	{\string\citation{#2}}\fi
\def\@citea{}\@cite{\@for\@citeb:=#2\do
	{\@citea\def\@citea{,}\@ifundefined
	{b@\@citeb}{{\bf ?}\@warning
	{Citation `\@citeb' on page \thepage \space undefined}}
	{\csname b@\@citeb\endcsname}}}{#1}}

\newif\if@cghi
\def\cite{\@cghitrue\@ifnextchar [{\@tempswatrue
	\@citex}{\@tempswafalse\@citex[]}}
\def\citelow{\@cghifalse\@ifnextchar [{\@tempswatrue
	\@citex}{\@tempswafalse\@citex[]}}
\def\@cite#1#2{{$\null^{#1}$\if@tempswa\typeout
	{IJCGA warning: optional citation argument 
	ignored: `#2'} \fi}}

\def\@refcitex[#1]#2{\if@filesw\immediate\write\@auxout
	{\string\citation{#2}}\fi
\def\@citea{}\@refcite{\@for\@citeb:=#2\do
	{\@citea\def\@citea{, }\@ifundefined
	{b@\@citeb}{{\bf ?}\@warning
	{Citation `\@citeb' on page \thepage \space undefined}}
	\hbox{\csname b@\@citeb\endcsname}}}{#1}}

\def\@refcite#1#2{{#1\if@tempswa\typeout
        {IJCGA warning: optional citation argument
	ignored: `#2'} \fi}}

\def\refcite{\@ifnextchar[{\@tempswatrue
	\@refcitex}{\@tempswafalse\@refcitex[]}}

\def\pmb#1{\setbox0=\hbox{#1}
	\kern-.025em\copy0\kern-\wd0
	\kern.05em\copy0\kern-\wd0
	\kern-.025em\raise.0433em\box0}

\def\fnt#1#2{\footnotetext{\kern-.3em
	{$^{\mbox{\scriptsize #1}}$}{#2}}}

\def\runninghead#1#2{\pagestyle{myheadings}
\markboth{{\protect\footnotesize\it{\quad #1}}\hfill}
{\hfill{\protect\footnotesize\it{#2\quad}}}}
\font\tenrm=cmr10
\font\tenit=cmti10 
\font\tenbf=cmbx10
\font\bfit=cmbxti10 at 10pt
\font\ninerm=cmr9
\font\nineit=cmti9

\font\eightrm=cmr8

\def\qed{\hbox{${\vcenter{\vbox{			%HOLLOW SQUARE
   \hrule height 0.4pt\hbox{\vrule width 0.4pt height 6pt
   \kern5pt\vrule width 0.4pt}\hrule height 0.4pt}}}$}}

\begin{document}

\runninghead{H.C. Rosu: Classical and quantum inertia}
%$\ldots$}
{H.C. Rosu: A matter of principles}
%$\ldots$}

%Comment (HCR): produce fraza de mai sus la inceputul fiecarei pagini

\normalsize\textlineskip
\thispagestyle{empty}
\setcounter{page}{1}

\copyrightheading{}
			%{Vol. 5, No.2(18) (June 1999) 81-91}

\vspace*{0.88truein}

%\fpage{1} %%%%%%%%%%%%%%%%%%%%%%%%%%%%%%%%%%%%%%%%%%%%%%%%%%%%%%%%%%%
\centerline{\bf CLASSICAL AND QUANTUM INERTIA: A MATTER OF PRINCIPLES}
\vspace*{0.035truein}
%\centerline{\bf MANUSCRIPTS USING COMPUTER SOFTWARE\footnote{For
%the title, try not to use more than 3 lines. Typeset the title
%in 10 pt Times Roman, uppercase and boldface.}}
\vspace*{0.37truein}
\centerline{\footnotesize HARET C. ROSU}
%\footnote{Typeset names in
%10 pt Times Roman, uppercase. Use the footnote to indicate the
%present or permanent address of the author.}}
\vspace*{0.015truein}
\centerline{\footnotesize\it Instituto de F\'{\i}sica,
Universidad de Guanajuato, Apdo Postal E-143, Le\'on, Gto, Mexico}
\baselineskip=10pt
%\centerline{\footnotesize\it City, State ZIP/Zone,
%Country\footnote{State completely without abbreviations, the
%affiliation and mailing address, including country. Typeset in 8
%pt Times Italic.}}
\vspace*{10pt}
%\centerline{\footnotesize SECOND AUTHOR}
%\vspace*{0.015truein}
%\centerline{\footnotesize\it Group, Laboratory, Address}
%\baselineskip=10pt
%\centerline{\footnotesize\it City, State ZIP/Zone, Country}
\vspace*{0.225truein}
%\publisher{(6 April 1999)}{(Date 2, Year)}

\vspace*{0.21truein}
\abstracts{
{\it ``But I, Simplicio, who have made the test can assure you that ..."}\\
Galileo Galilei,\cite{gaga}
}{}{}

\vspace*{0.13truein}
\abstracts{
A simple, general discussion of the problem of inertia is provided both
in classical physics and in the quantum world. After briefly reviewing
the classical principles of equivalence
(weak (WEP), Einstein (EEP), strong (SEP)), I pass to a presentation
of several equivalence statements in nonrelativistic quantum mechanics and
for quantum field vacuum states.
It is suggested that a reasonable type of preferred quantum field vacua may
be considered those possessing stationary spectra of their vacuum
fluctuations with respect to accelerated classical trajectories.
%PACS number(s):   04.20.Cv, 04.60.+n
}{}{}

%\vspace*{10pt}
%\keywords{The contents of the keywords}

\textlineskip                  %) USE THIS MEASUREMENT WHEN THERE IS
\vspace*{12pt}                 %) NO SECTION HEADING

\vspace*{1pt}\textlineskip	%) USE THIS MEASUREMENT WHEN THERE IS
%\section{General Appearance}    %) A SECTION HEADING
\vspace*{-0.5pt}
\noindent

\noindent
%%%%%%%%%%%%%%%%%%%%%%%%%%%%%%%%%%%%%%%%%%%%%%%%%%%%%%%%%%%%%%%%%%%%%

%\newpage

%\pagebreak

%\textheight=7.8truein
%\setcounter{footnote}{0}
%\renewcommand{\thefootnote}{\alph{footnote}}

%\section{The Main Text}
\noindent

%%%%%%%%%%%%%%%%%%%%%%%%%%%%%%%%%%%%%%%%%%%%%%%%%%%%%%%%%%%%%%%%%%%%%%%%%%%

%%%%%%%%%%%%%%%%%%%%%%%%%%%%%%%%%%%%%%%%%%%%%%%%%%%%%%%%%%%%%%%%%%%%%%%%%
\section{INTRODUCTION}
%%%%%%%%%%%%%%%%%%%%%%%%%%%%%%%%%%%%%%%%%%%%%%%%%%%%%%%%%%%%%%%%%%%%%%%%%%

Either empirical evidence or pure thought scientific belief (i.e., supported
by some mathematics)
can produce powerful physical principles for fundamental theories whenever an
appropriate interpretation is provided. This is the case of the
remarkable {\em universality}
of the {\em classical free fall} (Galileo's or the Pisa free fall;
first actual experiments in June 1710 at St. Paul's in London by Newton)
discovered at the
very beginning of modern science, and much later, but still before the
true advent of quantum mechanics, interpreted by Einstein
in terms of a universal coupling of all forms of matter to a common
metric tensor, an idea which was the key point in constructing
general relativity.
Indeed, since classical free fall motions (for the quantum case, see below),
although accelerating ones, do not depend on the test mass
($t_{f}=\sqrt{2h/g}$),
%which is the common parameter determining material inertia,
one may think of relating them
to fundamental predynamical (geometrical) properties of the universe.

On the other hand, by his second law, Newton imprinted on us
the idea that there is a profound relationship between material inertia
and all sorts of mechanical forces.
%perhaps the most simple connection between dynamics and kinematics.
Newton classified
forces into those of contact and those of action-at-a-distance. In the
first case, the agent producing the force is in direct contact with the test
particle (zero range forces), whereas in the latter the agent is able to
exert its effect instantaneously over huge distances (infinite range forces)
without any apparent transport by means of a {\em medium}.
Newton's second law
applies more to the contact forces and/or in situations in which
we can think of a close agent of forces, but of course the law is considered
as general.
The second category of forces
are the long range forces, which usually are tackled within electromagnetism
and gravitation. The discovery of a limit for the velocity of
signals (i.e., the velocity of light in vacuum, $c$) showed that
Newton's equation may well be substituted by
differential wave equations, including terms due to the limiting velocity
(i.e., operators of the form $-c^2\nabla ^2$). In this way, forces can be
transmitted at infinite speed only if the system is connected through
a `mass' term to an infinitely rigid substrate.\cite{ln}

Newton was always careful with the concept of mass, as he introduced it in
at least three of his basic formulas, which in fact referred to the same
concept of force. Since the accelerations look different in different
physical contexts,
Newton distinguished between inertial, passive
gravitational and active gravitational masses. A well-known review
of Bondi,\cite{bon} on these topics where the negative mass concept
is introduced is good reading. Later, the negative gravitational mass has
been restricted to antimatter by Morrison and Gold.\cite{mg}

If one focuses more on the concept of inertia
there occurs another difficulty, as we cannot be sure that this
apparently genuine feature of the test particle is determined by the
local conditions or
by some sort of global interaction with the whole rest of the universe,
a famous
alternative known as Mach's cosmological principle (for reviews and
connection with the notion of isotropic singularity, see Ref.[\refcite{t}]).
As is well known, already in 1710 Bishop Berkeley objected to
Newton's absolute space, insisting on the idea that it is not
meaningful from the experimental point of view,
thus being a forerunner of Einstein, who acted in the same way regarding
the mechanical ether.
In the spirit of the same idea that only systems in relative motion can
be detected, Mach attributed
inertial forces to acceleration relative to the ``heaven of fixed stars".
In more practical terms, Mach's principle of inertia (MPI) can be formulated
as follows,\cite{gran}

\vskip 0.1cm

{\bf MPI: more or less standard}

\noindent
{\sl The inertia of particles and bodies on earth and in the solar system is
due to their acceleration relative to all matter outside the solar system}.

\vskip 0.1cm

The idea of a cosmological scale of inertia was tackled
in interesting works
%belongs to  %has been discussed, among others,
by
Sciama,\cite{sc} still within the framework of general relativity,
%who made use of the Whitrow-Randall \cite{wr}
%cosmological relation $G\rho\sim t^{-2}$,
and culminated in the Brans-Dicke
scalar-tensor theory of gravitation.\cite{dick}

%One can argue
%that inertial properties as stated by Newton are never global while
%the Machian concept of inertia is never local, and hence they are
%completely opposite.
%But one could also work in the sense of
%mitigating the contrast between the two conceptions by specifying
%more carefully the conditions of passing from one to the other.

The main purpose of this paper is to bring together in one place works of
many people both on the classical inertia and the non-relativistic and
relativistic quantum features of this fundamental physical concept.
Several formulations of the principles of equivalence are reviewed and
possible connections and hints on further progress in this broad research
area are suggested.

\bigskip

%%%%%%%%%%%%%%%%%%%%%%%%%%%%%%%%%%%%%%%%%%%%%%%%%%%%%%%%%%%%%%%%%%%%%%%
\section{CLASSICAL EQUIVALENCES: WEP, EEP, SEP}
%%%%%%%%%%%%%%%%%%%%%%%%%%%%%%%%%%%%%%%%%%%%%%%%%%%%%%%%%%%%%%%%%%%%%%%%

Due to their very general/philosophical content the equivalence
assertions are subject to many contradictory opinions.
Will,\cite{w} has written excellent reviews
presenting the various formulations of the classical
principles of equivalence,
as well as their tests. In Will's works one can find some of the most general
statements of the classical inertia principles, which are quoted in the
following.
The assertions of Galilei and Newton are known at the present time as the
weak equivalence principle (WEP) and reads
%WEP states that

\vskip 0.3cm

{\bf WEP: Will}

\noindent
{\sl if an uncharged test body is placed at an initial
event in space-time and given an initial velocity there, then its
subsequent trajectory will be independent of its internal structure and
composition}.

\vskip 0.3cm

By uncharged test body one means an electrically neutral
body that has negligible self-gravitational energy and moreover is small
enough in size in order to neglect the coupling to the inhomogenities of
the external fields. By means of modern E\"otv\"os-type experiments
(Roll, Krotkov and Dicke,\cite{e1} Braginsky and Panov,\cite{e2}
Adelberg,\cite{e3})
the WEP is clearly correct at a fractional precision better
than 10$^{-11}$.

What is known as the Einstein equivalence principle (EEP) is the following
statement

\vskip 0.3cm

{\bf EEP: Will}

\noindent
{\sl the outcome of any local non-gravitational test experiment is
independent of where and when in the universe it is performed}.

\vskip 0.3cm

By a local non-gravitational test
experiment one should mean any experiment that is performed in a freely
falling laboratory which is small and shielded with respect to the
inhomogeneities in the external fields and for which self-gravitational
effects are negligible.
If EEP is valid, then gravitation must be a curved space-time phenomenon,
and therefore one can think of metric theories of gravity. This
is the reason why EEP is so important. The EEP as stated by Will is perhaps
an excessive general phrase.
Following the work of Kreinovich and Zapatrin,\cite{kz}
I quote the standard EEP (i.e., almost as given by Einstein)

\vskip 0.3cm

{\bf EEP: standard}

\noindent
{\sl What ever measurements we perform inside some spacetime region we cannot
distinguish between the case when there is a homogeneous gravitational
field and the case when all bodies in this region have constant acceleration
with respect to some inertial frame. And since any field can be considered
homogeneous in a small enough region, the principle can be applied to a
neighborhood of any point}.

\vskip 0.3cm

V.A. Fock,\cite{fock}
pointed out that this formulation is not exact
enough, because in the presence of gravity the spacetime curvature tensor is
nonzero, while in a uniformly accelerated frame the curvature tensor is
zero.

The standard EEP shows plainly the correctness of Ohanian's opinion,\cite{oh}:
``Einstein's theory of general relativity was conceived
in an attempt at formulating a relativity of acceleration".

%The general topic
%of {\em acceleration} is in an active period of opinions \cite{acc}.

The most general equivalence formulation
appears to be the strong equivalence principle (SEP) dealing
with situations in which one considers in addition to the
metric field other types of dynamical fields and/or prior-geometric
 fields.
The hypothesis is that all these fields yield local gravitational
physics which may have both location and velocity-dependent effects.
SEP states that

\vskip 0.3cm

{\bf SEP: Will}

\noindent
{\sl the outcome of any local test experiment is
independent of the velocity of the freely falling apparatus and of where
and when in the universe it is performed}.

\vskip 0.3cm

The distinction between
SEP and EEP is the inclusion of bodies with self-gravitational
interactions (planets, stars, black-holes). Actually, SEP means the equality
of the so-called passive gravitational mass and inertial mass.
If SEP is valid, there must be one and only one
gravitational field in the universe given by a unique metric.
Laboratory experiments test only WEP in the form of composition
dependent interactions of Yukawa form. Tests of SEP, on the other hand,
are only possible in astrophysical environments, first of all in the
Solar System, since one needs bodies with a significant contribution
to their inertia from their gravitational binding energy.
A useful parameter $\eta$ measuring the deviations from SEP has been
introduced by Nordtvedt,\cite{Nor} in 1968. In other
words, SEP means to determine if gravitational binding energies are
falling with the same acceleration. The predilect test of SEP is lunar laser
ranging (i.e., the change in round-trip radar time between Earth and Moon
when the radar path passes close to the Sun).
The data show that the fractional difference in the falling
accelerations toward the Sun between the (iron-dominated) Earth and the
(silica-dominated) Moon is $(2.7\pm 6.2)\times 10^{-13}$.\cite{dic}
This is still a weak binding case since the gravitational binding energy
reduces the mass of the Earth by only 5.1 parts in $10^{10}$. More recently,
Earth-Mars ranging was proposed,\cite{mars} as well as strong field
regimes.\cite{wex}

As a matter of fact, in all forms of SEP the weak point is the
notion of locality. This problem has been tackled in the important paper of
Bertotti and Grishchuk.\cite{bg} Usually, the locality concept required
by equivalence
is to say that the effects of curvature on the local metric are negligible.
According to Bertotti and Grishchuk, in an
appropriate inertial frame and in the slow-motion approximation
{\it a local gravitational system} can be defined whenever the measurement
errors are greater than the corresponding effects of tidal forces.
The formulation of SEP belonging to these authors is the following

\vskip 0.3cm

{\bf SEP: Bertotti and Grishchuk}

\noindent
{\sl We say that SEP is fulfilled if, when the size r of the system
is sufficiently small, its dynamical behaviour, to a given accuracy, is
universal and not affected by the external world}.

\vskip 0.3cm

Moreover, these authors discuss the problem of the universality
of {\em gravitational clocks}, commenting on the three cases previously
considered in the works of Will,\cite{w1} namely the
{\em rotating relativistic star},
the {\em slowly rotating black hole} and the {\em binary system}.
In Newtonian gravity the
gravitational (Kepler) clocks can be considered {\em universal},
if their size is small enough. This comes out when one takes into
account the effects of a third body on the two-body system. Will has found
in all the three cases the common changes in the frequency due to special
relativity and the gravitational shift formula.
The problem is
to estimate the changes in the laws of gravitational
two-body systems when the relativistic corrections are included in order to see
to what approximation they can be considered universal.
On the other hand, there is considerable technological interest in the
development of highly stable spaceborne clocks that may lead to the detection
of the gravitomagnetic field of the earth according to the
recent proposal of Mashhoon and collaborators.\cite{mash}
%Indeed, it is known that the differentials
%of the relativistic transformations are nonlinear \cite{ng}, and
%consequently in high gravitational fields the two-body problem may change
%drastically.
%The universal behavior will be affected by tidal forces and
%the nonlinear interactions with the rest of the world.
%In general, the SEP conditions of weak-field and slow-motion are
%satisfied only within PPN theories.

%Very interesting would be to study the clock universality problem at a
%completely different scale, namely
%in the case of classical-like Rydberg wave packets in atoms and antimatter
%atoms
%(especially for the forthcoming laboratory antihydrogen),
%and to see in
%detail what would be the significance of a Rydberg universal clock in this
%context. A paper by Pinto \cite{pin} and literature on the equivalence
%principle for antimatter \cite{huh} are to be consulted in this respect.

Coming back to WEP, one should mention that various authors put it at the
ground of
detailed theoretical constructions leading to nice mathematical consequences.
One of the best known procedures is the Ehlers-Pirani-Schild
scheme.\cite{eps} These authors argue that WEP has two important
features from the theoretical point of view. For a space-time
manifold with a pure gravitational field, one can say that
(a) the possible motions of all
freely falling test particles are the same, and
(b) at any point $p$ in space-time, there exists a neighborhood $U(p)$ of
$p$ and a four-dimensional coordinate system, such that the trajectories of
{\em every} freely falling test particle through $p$ satisfies
$d^2x^{\mu}/d\lambda ^{2}=0$ at $p$ for a suitable parameter $\lambda$ along
the trajectory. This is just a local form of the law of inertia, and such a
coordinate system is said to be locally inertial at $p$. The latter condition
is a property shared only by gravitational fields, which being
related to the connection coefficients, can be
coordinate transformed away. Using this special property for the massive and
massless cases, it was shown by Ehlers, Pirani and Schild, that there exists
an affine connection $\omega$, independent of the test particles used,
such that the trajectories of freely falling test
particles are affinely parametrized geodesics with respect to it.

\bigskip

%%%%%%%%%%%%%%%%%%%%%%%%%%%%%%%%%%%%%%%%%%%%%%%%%%%%%%%%%%%%%%%%%%%%%%%%%%%
\section{INERTIA PRINCIPLES IN THE QUANTUM WORLD}
%%%%%%%%%%%%%%%%%%%%%%%%%%%%%%%%%%%%%%%%%%%%%%%%%%%%%%%%%%%%%%%%%%%%%%%%%%

Two fundamental concepts of quantum theories are the intrinsic vacuum noise
known as zero point energy and the quantum state (although one may prefer
the path-integral formalism).
The classical concept of trajectory occurring in the classical formulations
of the inertia principles can be considered a sort of
zero-order approximation at the best, and
the space-time picture is only one of the many possible representations.
Moreover, time enters the quantum formalism merely as a parameter and it is
hard if not impossible to think of a time operator as happens for other
common observables. Many different concepts of quantum time and/or clocks
are actively pursued. I recall here only the {\em optimal quantum clock}
concept of Bu\v zek, Derka and Massar,\cite{qclo} based on trapped ions,
the {\em tunneling times},\cite{csopt} and
the {\em flavor-oscillation clock}.\cite{ahc}
Probabilistic arguments are practically
unavoidable when discussing scales at and below the molecular ones, and
therefore the quantum equivalence statements are expected to
be substantially different from their classical counterparts. Moreover,
considering quantum mechanics as a sort of wave theory,
the mass parameter will manifest itself at the experimental level mainly
through de Broglie, Compton, and Planck (wave)lengths, i.e.,
in (non-relativistic) matter interferometry,\cite{gb,lam}
particle production (one can look at some bremsstrahlung literature),
and `gravity-wave' interferometry,\cite{gac} experiments.

\bigskip
%\bigskip

{\bf 3.1 Nonrelativistic quantum inertia}
%%%%%%%%%%

For the nonrelativistic quantum mechanics, one can find
interesting aspects of the equivalence principles in various
theoretical (see e.g., Ref.[\refcite{ph}]) and
experimental approaches, especially those
related to neutron,\cite{gb} and atom,\cite{lam}
beam interferometry.

The first quantum experiments in the EP context have been of
Galilei type (Pisa gedanken experiment)
and have been performed with neutrons,\cite{dabbs} according to the
philosophy ``let's see how they fall !", although the free fall of atoms in
molecular beams was easily observable since late thirties, but was used for
other purposes, such as to measure the Bohr magneton by compensating
gravity through magnetic fields.\cite{s37} The WEP has been confirmed for
neutrons
to within 3$\cdot$ 10$^{-3}$ accuracy, by measuring the fall height of a
neutron, initially moving horizontally at a known velocity. At present,
there are hopes that experiments with ultracold neutrons,\cite{pok} can
lead to an accuracy as high as $10^{-6}$.
Many other proposals have been made over the years,
such as antimatter in free fall,\cite{f}
cooled atoms in optical molasses,\cite{mol} and opto-gravitational
cavities,\cite{cav} and recently free-falling
mesoscopic Schr\"odinger cat states,\cite{cat} and, possibly, atomic Bose
condensates.

Recently, Ahluwalia,\cite{flav} revived an argument provided by
Kenyon,\cite{ken} on the possible observability of constant gravitational
potentials by
gravitationally induced CP violation this time in the context of earth
and/or solar-bound {\em quantum-mechanical free falls}. Using the simplest
example of a linear superposition of two different mass eigenstates, which in
neutrino physics led him and Burgard to the concept of flavor-oscillation
clocks,\cite{ahc} Ahluwalia argues
that the redshift-inducing phases of such freely-falling clocks depend
directly on the extremely small constant gravitational potential of the local
cluster of galaxies, the so-called Great Attractor.
It will be interesting to investigate Ahluwalia's suggestion using
the wave packet representation,\cite{nau} for both meson and
neutrino oscillations. Furthermore, it is known that Kenyon's paper was
criticized by
Nieto and Goldman in their classic 1991 {\em Physics Report}, according to
the canonical opinion that no independent experimental means are available
to measure absolute gravitational potentials. However, as pointed out by
Ahluwalia,\cite{circum} the weak field limits of classical gravity and
{\em any} theory of quantum gravity have different behavior with respect
to the gravitational potential. Thus, the possibility of such a Aharonov-Bohm
type situation in quantum free-fall remains open. Also, according to
Ahluwalia, for the case of quantum freely falling frames
there is the possibility of
violation of the {\em local position invariance},\cite{lpi} which together
with the {\em local Lorentz invariance},\cite{lLi} stands at the
ground of EEP.

In the context of matter wave interferometry,
L\"ammerzahl,\cite{lam} proposed a generalized quantum WEP (QWEP), which
reads as follows

\vskip 0.1cm
%\newpage

{\bf QWEP: L\"ammerzahl}

\noindent
{\sl For all given initial states the input independent result of a
physical experiment is independent of the characteristic parameters (like
mass, charge) of the quantum system}.

L\"ammerzahl carefully added a few more comments claryfing
the notions he was using in the above statement, and proceeded to show that
some important classes of quantum quantities, like
the gravity-induced phase shifts of atom beams and neutron interference
experiments and the time evolution of expectation values and uncertainties
support his formulation.

%The equivalence considerations are very interesting in quantum field theories,
%and, as a matter of fact, my main motivation in writing this paper was to

\bigskip
%\bigskip

{\bf 3.2 Relativistic field inertia}
%%%%%%%%%%

There are deep insights in the problem of quantum field inertia that
have been gained in the last two decades as a consequence of Hawking effect
and Unruh effect, and consequently the `imprints' of gravitation in the
relativistic quantum physics have been substantially clarified.\cite{ug}
The method of quantum detectors (accelerated elementary
particles) proved to be very useful for the understanding of
the quantum field inertial features. A new way of thinking of
quantum fluctuations has emerged and new
pictures of the vacuum states have been provided, of which the landmark one is
the heat bath interpretation of the Minkowski vacuum state from
the point of view of a uniformy accelerating non-inertial quantum detector.
This interpretation is
mostly attributed to Unruh because of his 1976 seminal paper, although the
corresponding mathematical formula has been obtained more or less at the
same time by several people.\cite{dav} One can think of
{\em zero-point fluctuations}, {\em gravitation} and {\em inertia} as the
only three universal phenomena of nature. This idea has been popularized
by Smolin,\cite{smo}
some time ago. However, one may also think of inertia as related to a
peculiar sort of collective degrees of freedom known as vacuum
expectation values (vev's) of Higgs fields. As we know,
these vev's do not follow from the fundamentals of quantum theory.
On the other hand, one can find
papers claiming that inertia can be assigned to a Lorentz type force
generated by electromagnetic zero-point fields,\cite{hrp}
It is also quite well known the
{\em Rindler condensate} concept of Gerlach.\cite{ger} The point is that
there exist completely coherent zero-point condensates
(like the Rindler-Gerlach one) entirely mimicking the Planck spectrum, without
any renormalization (Casimir effect).

According to Unruh,\cite{u} simple model particles of uniform,
one-dimensional proper acceleration $a$ in Minkowski vacuum are immersed
in a scalar quantum field `heat' bath of temperature
$$
T_{a}=\frac{\hbar}{2\pi c k}\cdot a~,
$$
where $\hbar$ is Planck's constant, $c$ is the light speed in
vacuum, and $k$ is Boltzmann's constant. For first order corrections to this
formula see works by Reznik.\cite{rez}
The Unruh temperature is
proportional to the lineal uniform acceleration, and the scale of such
noninertial quantum field `heat' effects is
fixed by the numerical values of universal constants to the very low value of
$4\times 10^{-23}$ in cgs units). In other words, the huge acceleration of
$2.5 \times 10^{22}$ ${\rm cm/s}^2$ can produce a black body spectrum of
only 1 K.
In the case of Schwarzschild black holes, using the surface gravity
$\kappa=c^{4}/4GM$ instead of $a$, one gets the formula for their
Hawking temperature, $T_{\kappa}$.
In a more physical picture, the Unruh quantum field heat reservoir is
filled with the so-called Rindler photons
(Rindler quasi-particles), and
therefore the quantum transitions are to be described as absorptions or
emissions of the Rindler reservoir photons.
The Unruh picture can be used for interpreting
Hawking radiation in Minkowski space.\cite{jac} In order to do that,
one has to consider the generalization(s) of the equivalence principle to
quantum field processes. A number of authors have discussed this important
issue with various degrees of detail and meaning and with some debate.\cite{1}

Nikishov and Ritus,\cite{nr} raised the following objection to the
heat bath concept. Since absorption and emission processes
take place in finite space time regions, the application of the
local principle of equivalence requires a constant acceleration
over those regions. However, the space-time extension of the quantum
processes are in general of the order of inverse acceleration. In Minkowski
space it is not possible to create homogeneous and uniform
gravitational fields having accelerations of the order of $a$ in
spacetime domains of the order of the inverse of $a$.   %$a^{-1}$.

Pinto-Neto and Svaiter,\cite{pns} summarized the detailed discussions
of Grishchuk, Zel'dovich, and Rozhanskii, and of
Ginzburg and Frolov, concerning the formulations of quantum field
equivalence principles from the point of view of the response functions of
quantum detectors, in particular the Unruh-DeWitt (UDW) two-level monopole
detector.
Recall
that in the asymptotic limit the response function is the integral of the
quantum noise power spectrum. Or, since the derivative of the response
function is the quantum transition rate, the latter is just the measure of
the vacuum power spectrum along the chosen trajectory (worldline) and in
the chosen initial (vacuum) state. This is only in the asymptotic limit
and there are cases requiring calculations in finite time intervals.\cite{ss}
Denoting by $R_{M,I}$, $R_{R,A}$, and $R_{M,A}$ the detection
rates with the first subscript corresponding to the vacuum (either Minkowski
or Rindler) and the second subscript corresponding to either inertial or
acceleratig worldline, one can find for the UDW detector in a
scalar vacuum that $R_{M,I}=R_{R,A}$ expressing the dissipationless
character of the vacuum fluctuations in this case, and a thermal factor
for $R_{M,A}$ leading to the Unruh heat bath concept. In the case of
a uniform gravitational field, the candidates for the vacuum state are
the Hartle-Hawking ($HH$) and the Boulware ($B$) vacua. The $HH$ vacuum is
defined by choosing incoming modes to be positive frequency modes with respect
to the null coordinate on the future horizon and outgoing modes as
positive frequency ones with respect to the null coordinate on the past
horizon, whereas the $B$ vacuum has the positive frequency modes with
respect to the Killing vector which makes the exterior region static.
For a uniform gravitational field the $HH$ vacuum can be thought of as the
counterpart of the Minkowski vacuum, while the $B$ vacuum is the equivalent
of the Rindler vacuum. Then, the quantum field equivalence principle
(QFEP) can be formulated in one of the following ways

\vskip 0.3cm

%\underline{
{\bf Quantum detector-QFEP: $HH-M$ equivalence}
%}

\noindent
{\sl i) The detection rate of a free-falling UDW detector in the
HH vacuum is the same as that of an inertial UDW detector in the M vacuum.

\noindent
ii) A UDW detector at rest in the HH vacuum has the same DR as a uniformly
accelerated detector in the M vacuum.}

\vskip 0.3cm

%\underline{
{\bf Quantum detector-QFEP: $B-R$ equivalence}
%}

\noindent
{\sl iii) A UDW detector at rest in the B vacuum has the same detection rate
as a uniformly accelerated detector in the R vacuum.

\noindent
iv) A free-falling UDW detector in the B vacuum has the same detection rate
as an inertial detector in the R vacuum.}

\vskip 0.3cm

The above formulations seem reasonable enough, but their generalization
to more realistic cases must be carefully considered in the future.
Let us record one more formulation due to Kolbenstvedt,\cite{1}

\vskip 0.3cm

{\bf Quantum detector-QFEP: Kolbenstvedt}

\noindent
{\sl A detector in a gravitational field and an accelerated detector will
behave in the same manner if they feel equal forces and perceive radiation
baths of identical temperature}.

\vskip 0.1cm

In principle, since the Planck spectrum is Lorentz invariant
(and even conformal invariant) its inclusion in equivalence statements
looks quite natural. The linear connection between temperature and
one-dimensional, uniform, proper acceleration, which is also
valid in some important gravitational contexts, is indeed a
fundamental relationship, because it allows for an absolute meaning of
quantum field effects in such an {\em ideal} noninertial frame, as soon as
one recognize thermodynamic temperature as
the only {\it absolute}, i.e., fully {\it universal} energy type physical
concept.
%However, the situation is not as simple as that.
In general, the scalar quantum field vacua
are not stationary stochastic processes (stationary vacuum noises)
for all types of classical trajectories. Nevertheless, the lineal
acceleration is {\em not} the only case with that property as was shown by
Letaw,\cite{let} who extended Unruh's considerations obtaining six types
of worldlines with
stationary vacuum excitation spectrum (SVES-1 to SVES-6, see below),
as solutions of some generalized Frenet equations
and under the condition of constant curvature invariants of the worldline
(curvature, torsion and hypertorsion, i.e., $\kappa$, $\tau$ and $\nu$,
respectively). The six stationary cases are the following

%\noindent
1. $\kappa =\tau=\nu=0$, (inertial worldlines; trivial cubic SVES-1).

%\noindent
2. $\kappa \neq 0$, $\tau=\nu=0$,
(hyperbolic worldlines; SVES-2 is Planckian allowing the
interpretation of $\kappa/2\pi$ as `thermodynamic' temperature).

%\noindent
3. $|\kappa|<|\tau|$, $\nu=0$, $\rho ^2=\tau ^2-\kappa ^2$,
(helical worldlines; SVES-3 is an analytic function
corresponding to case 4 below only in the limit $\kappa\gg \rho$).

%\noindent
4. $\kappa=\tau$, $\nu=0$,
(the spatially projected worldlines are semicubical parabolas containing a
cusp where the direction of motion is reversed; SVES-4 is analytic in the
dimensionless energy variable involving $\kappa$, but is not Planckian).

%\noindent
5.  $|\kappa|>|\tau|$, $\nu=0$, $\sigma ^2=\kappa ^2-\tau ^2$,
(the spatially projected worldlines are catenaries; SVES-5 cannot be found
analitically in general, but for $\tau/\sigma\rightarrow 0$ tends to become
Planckian (SVES-2), whereas for $\tau/\sigma\rightarrow\infty$ tends toward
SVES-4).

%\noindent
6. $\nu\neq 0$,
(rotating worldlines uniformly accelerated normal to their plane of rotation;
SVES-6 forms a two-parameter set of curves).

As one can see only the hyperbolic worldlines allow for a Planckian SVES
and actually for a one-to-one
mapping between the curvature invariant $\kappa$ and the `thermodynamic'
temperature.
Thus, one can infer that in some cases it is possible to determine the
classical worldline on which a quantum particle is moving from measurements
of the vacuum noise spectrum. There is much interest in considering the
radiation patterns at accelerators in this perspective,\cite{mont} and
it is in this sense that a sufficiently general and acceptable statement
on the {\em universal} nature of
the kinematical parameters occurring in a few important quantum field model
problems can be formulated as follows

\vskip 0.1cm

\noindent
{\em  There exist accelerating
classical trajectories (worldlines) on which moving ideal (two-level)
quantum systems can detect the scalar vacuum environment as a stationary
quantum field vacuum noise with a spectrum directly related to
the curvature invariants of the worldline, thus allowing for a
radiometric meaning of those invariants}.

\vskip 0.1cm

Another important byproduct is the possibility to choose a class of
preferred vacua of the quantum world,\cite{pr} as {\em all} those
having stationary vacuum noises with
respect to the classical (geometric) worldlines of constant curvature
invariants because in this case one may find some
necessary attributes of universality in the more general
quantum field radiometric sense,\cite{rad} including as a particular case the
Planckian thermal spectrum.
%(Note: there is
%an interesting fingerprint of the number of dimensions discovered by
%Shin Takagi).
Of course, much work remains to be done towards a more ``experimental"
picture of highly academic calculations in quantum field theory, which are
to be considered as useful only as a guide for more definite
and therefore more complex situations.\cite{mont}
A careful look to the literature
shows that there are already steps in this direction. For example,
Nagatsuka and Takagi,\cite{nt}
studied radiation from a quasi-uniformly accelerated charge;
Cresser,\cite{cres} considered a
model electron detector similar to the DeWitt monopole detector allowing
him to develop a theory of electron detection and
photon-photoelectron correlations in two-photon ionization processes;
Klyshko,\cite{kly} discussed the possible connection between
photodetection, squeezed states and accelerated detectors;
Frolov and Ginzburg,\cite{fg} pointed out that the radiation associated
to uniformly accelerated detectors moving in vacuum is similar to that
occurrring in the region of anomalous Doppler effect, which take place when
a quantum detector is moving at a constant superlight velocity in a medium.
Marzlin and Audretsch,\cite{maau} considered constantly accelerated
multi-level atoms and concluded that the magnitude of
the Unruh effect is not modified.
However, one should notice that all the
aforementioned quantum field vacua look extremely ideal from the experimental
standpoint. Indeed, it is known that only strong external fields can make
the electrodynamical vacuum to react and show its physical properties,
becoming similar to a magnetized and polarized medium,
and only by such means one can learn about the structure of QED vacuum.
Important, recent results on the relationships between Schwinger mechanism
and Unruh effect have been reported in recent works.\cite{gab}

At the axiomatic level, Hessling,\cite{hes} published new results
on the algebraic quantum field equivalence principle (AQFEP).
Hessling's formulation is too technical to be reproduced here.
The difficulties are related to the rigorous formulation
of local position invariance, a requisite of equivalence, for the
singular short-distance behavior of quantum fields, and to the generalization
to interacting field theories.
%despite old results of Unruh and Weiss \cite{uw}.
Various general statements of locality,\cite{haa} for linear quantum fields
are important steps toward proper formulations of AQFEP. These are nice but
technical results coming out mainly from
clear mathematical exposition involving the Kubo-Martin-Schwinger
(KMS) states of Hadamard type.
Hessling's AQFEP formulation is based on the notion of quantum states
{\em constant up
to first order} in an arbitrary spacetime point, and means that for these
states a certain
scaling limit should exist, and moreover a null-derivative condition with
respect to a local inertial system around that arbitrary point is to be
fulfilled for all n-point functions. In a certain sense this is similar to
the properties of Gaussian noise. For example, the vacuum state of the
Klein-Gordon field in Minkowski space with a suitable scaling function
fulfilles Hessling's AQFEP.
Hessling showed, using as a toy model the asymptotically free $\phi ^3$ theory
in six-dimensional
Minkowski space, that the derivative condition of his QEP is not satisfied
by this interacting quantum field theory, which perturbatively is similar to
QCD. This failing is due to the running coupling constant which does not go
smoothly to zero in the short-distance limit.
The complexity of the Yang-Mills vacuum (YMV) is noteworthy.\cite{shif}
Interestingly,
Reuter and Wetterich,\cite{rw} claim that the true YMV is characterized
by a nonvanishing gluon condensate.
If so, one may think of a gluon-vev inertial contribution of the YMV.
Also, the ground state of quantum gravity,\cite{gsqg} although a highly
speculative topic might allow considerations from the inertia standpoint.

Finally, it is worth mentioning that the {\em time-thermodynamics} relation
in general covariant theories and the connection with Unruh's temperature and
Hawking radiation is an active area of research due to the remarkable
correspondence between causality and the modular Tomita-Takesaki
theory.\cite{crov} It would be interesting to formulate in this context some
sort of AQFEP statement beyond that of Hessling.

\bigskip
%\newpage

%%%%%%%%%%%%%%%%%%%%%%%%%%%%%%%%%%%%%%%%%%%%%%%%%%%%%%%%%%%%%%%%%%%%%%%%%%
{\bf 3.3 Brief miscellany}
%%%%%%%%%%%%%%%%%%%%%%%%%%%%%%%%%%%%%%%%%%%%%%%%%%%%%%%%%%%%%%%%%%%%%%%%%

This subsection is a browsing through further literature.

(i)
A discussion involving EEP in the Schwarzschild geometry has been recently
made by Moreau, Neutze, and Ross.\cite{mnr} They found a coordinate
transformation separating the line element into a pure acceleration,
diagonal part and an off-diagonal, pure curvature contribution allowing for
a good understanding of the equivalence issue in that case.

(ii)
Punsly,\cite{p} deals with the problem of equivalence as related to
black hole evaporation adopting the premises that at
each point in spacetime all the inertial observers can accurately
postulate the relativistic quantum field theories of flat spacetime
on open sets with dimensions much less than the radii of curvature
of spacetime. Thus, by pure local considerations all the local observers
can formulate number representations of the field through local
particle creation and annihilation operators. Each local freely falling
observer transports with him his own definition of the vacuum state.
The proposal of Punsly is to introduce a global vacuum state defined
throughout the spacetime outside of the horizon by ``integrating"
the local vacua along a space-filling family of freely falling
trajectories. Such a proposal satisfies the principle of equivalence.

(iii)
Kleinert,\cite{kl}
speaks about a new QEP which determines short-time action and
measure of fluctuating orbits in spaces with curvature and torsion that he
gets from a simple mapping procedure by which classical orbits and path
integrals for the motion of a point particle in flat space can be
transformed directly into those in curved space with torsion.
Alvarez and Mann,\cite{alma} evaluated the constraints on non-metric
violations of the EEP over a wide sector of the electroweak standard model
of particle physics.
On the other hand, Kauffmann,\cite{kff} tackled a gravity-induced
birefringence of space due to nonmetric coupling between gravity and
electromagnetism. He got an upper bound from time delay data of the
pulsar PSR 1937+21.

(iv)
Anandan,\cite{an}
introduces the concept of quantum physical geometry by applying the
Ehlers-Pirani-Schild scheme to freely falling quantum wavepackets.

(v)
Jaekel and Reynaud,\cite{paris} discuss for the case of a Fabry-Perot cavity
the radiation pressure due to
quantum fluctuations, which induces mechanical effects on scatterers.
They calculate the correction to the total mass of the cavity
resulting from the Casimir force between the two mirrors, and show that
energy stored in the vacuum fluctuations contributes to inertia in
conformity with the law of inertia of energy. It comes out that
inertial masses exhibit quantum fluctuations with a characteristic
mass noise spectrum. For a recent review of Casimir effect(s), see
Ref.[\refcite{kard}].

(vi)
There is a conjecture due to Gr\o n and Eriksen,\cite{ge} that Einstein's
field equations do not permit the existence of empty space-times with a
uniform gravitational field, i.e., a field in which the proper acceleration of
a free particle instantaneously at rest is the same everywhere in
the field. Apparently, as in electromagnetism, the closest one can come to a
uniform gravitational field in empty four-dimensional space-time is the
parallel gravitational field outside a massive plane of infinite extension.
%In the literature one can encounter interesting discussions concerning the
%relationships between the uniform accelerated rigid reference frames in
%Minkowski spacetime (as expressed in M\o ller coordinates) and the rigid
%frames in a uniform gravitational field.

(vii)
Carlini and Greensite,\cite{cg} show that the classical field equations of
general relativity can be expressed as a single geodesic equation, describing
the free fall of a point particle in superspace, and applied the result to
several minisuperspace cosmological models.

(viii) I also mention a work of
%i) Newton-Gauss principle states that the potential of an electrostatic or
%gravistatic field created by a finite spherically symmetric charge
%or mass density distribution is similar outside the source region
%to the field created by the whole charge or mass located at the central point
%of the charge or mass distribution. Recently, the Newton-Gauss principle
%has been generalized to wave equations \cite{kz}. This is
%of relevance for the discussion of inertia.
%ii)
Baumann and collaborators,\cite{bau} who reported highly interesting
experiments regarding the free fall
of immiscible vortex rings in liquids, which might have an impact on WEP,
if WEP oriented.

(ix)
In a series of papers of Faraggi and Matone,\cite{fm} a sort of mathematical
equivalence postulate is introduced stating that all physical systems can
be connected by a coordinate transformation to the free system with
vanishing energy, uniquely leading to the quantum analogue of the
Hamilton-Jacobi equation, which is a third-order non-linear differential
equation. By this means a trajectory representation of the quantum mechanics
is derived, depending on the Planck length.

(x)
When the vacuum noises are not stationary,
one can nevertheless perform their tomographical processing,\cite{manko}
requiring joint time and frequency information.
Alternatively,
since in the quantum detector method the vacuum autocorrelation functions
are the essential physical quantities, and since according to
fluctuation-dissipation theorem(s) (FDT) they are related to the linear
(equilibrium) response functions to an initial condition/vacuum,
more FDT type work, especially their generalization to the out of
equilibrium case,\cite{out} will be useful in this framework.
In fact, there is some recent progress due to
Hu and Matacz,\cite{hm} in making more definite
use of FDT for vacuum fluctuations.
Very recently, Gour and Sriramkumar,\cite{gs} questioned if small particles
exhibit Brownian motion in the quantum vacuum and concluded that even though
the answer is in principle positive the effect is extremely small and thus
very difficult to detect experimentally.
For the well-known method of influence
functionals and generalizations, see Ref.[\refcite{ang}].

\bigskip
 
%%%%%%%%%%%%%%%%%%%%%%%%%%%%%%%%%%%%%%%%%%%%%%%%%%%%%%%%%%%%%%%%%%%%%%%%%%
\section{CONCLUSION}
%%%%%%%%%%%%%%%%%%%%%%%%%%%%%%%%%%%%%%%%%%%%%%%%%%%%%%%%%%%%%%%%%%%%%%%%%%%

As I have presented here in some detail, considerations of equivalence type
in quantum field theories may well guide the abstract research towards the
highly required feature of universality (going up to the act of
measurement itself, see Ref.[\refcite{ume}]) which is one of the ultimate
purposes of the meaningful research.

The equivalence principles are related to a number of fundamental problems
that may be considered as ever-open-issues, like those
of physical {\em mass},\cite{mass}
%(usually, principle-of-equivalence reasoning is applied in
%tests done to put bounds on Yukawa type masses/interactions \cite{myuk}),
{\em locality},\cite{loc} and more rigorous definitions of {\em reference
frames},\cite{frame} either in
classical physics or in the quantum approach.
Indeed, referring to the latter issue, since the equivalence principles are
connected to the type of geometrical structure of spacetime(s), more
geometrical-axiomatic formulations,\cite{col} of such fundamental statements
are required for example in the context of fractal geometry,\cite{fgeom}
and even beyond geometry,\cite{beyond} in order to
learn for example under what conditions one can get a unique metric.

%Another important task would be to
%understand the inertial features of the quantum chromodynamical vacuum, and of
%other interacting field theories \cite{my}.

To this end, one can say that
as any matter of interpretation at a very general level,
the principles of equivalence are open to
many opinions and discussions. They have been the beginning of modern physics,
and probably they will ever frustrate us in one way or another.

\nonumsection{Acknowledgements}
\noindent
%%%%%%%%%%%%%%%%%%%%%%%%%%%%%%%%%%%%%%%%%%%%%%%%%%%%%%%%%%%%%%%%%%%%%%
This work was supported in part by the (Mexican) CONACyT
project 458100-5-25844E.
The author wishes to acknowledge
Drs. D.V. Ahluwalia, J. Socorro and V.I. Tkach
for their suggestions.

%\newpage
%%%%%%%%%%%%%%%%%%%%%%%%%%%%%%%%%%%%%%%%%%%%%%%%%%%%%%%%%%%%%%%%%%%%%%%%

%This section should come before the References. Funding
%information may also be included here.

\nonumsection{References}
%\noindent
%References are to be listed in the order cited in the text. Use
%the style shown in the following examples. For journal names,
%use the standard abbreviations. Typeset references in 9 pt Times
%Roman.

%\begin{enumerate}

%\appendix

%\noindent
%Appendices should be used only when absolutely necessary. They
%should come after the References. If there is more than one
%appendix, number them alphabetically. Number displayed equations
%occurring in the Appendix in this way, e.g.~(\ref{that}), (A.2),
%etc.
%\begin{equation}
%\mu(n, t) = {\sum^\infty_{i=1} 1(d_i < t, N(d_i) = n) \over
%\int^t_{\sigma=0} 1(N(\sigma) = n)d\sigma}\,. \label{that}
%\end{equation}
\end{document}